\documentclass[aps,amsmath,amssymb,twocolumn]{revtex4}
\newcommand{\dket}[1]{| \, #1 \rangle\!\rangle}
\newcommand{\dbra}[1]{\langle\!\langle #1 \, |}
\begin{document}
\title{Quantum universal detectors} \author{G. M.  D'Ariano, P. Perinotti, and
  M. F. Sacchi} \affiliation{Quantum Optics \& Information Group,
  Unit\`a INFM and Dipartimento di Fisica ``A. Volta'', Universit\`a
  di Pavia, via A. Bassi 6, I-27100 Pavia, Italy}
\homepage{http://www.qubit.it} 
\begin{abstract}
  We address the problem of estimating the expectation value $\langle
  O\rangle$ of an arbitrary operator $O$ via a universal measuring
  apparatus that is independent of $O$, and for which the expectation
  values for different operators are obtained by changing only the
  data-processing. The ``universal detector'' performs a joint
  measurement on the system and on a suitably prepared ancilla. We
  characterize such universal detectors, and show how they can be
  obtained either via Bell measurements or via local measurements and
  classical communication between system and ancilla.
\end{abstract}
\maketitle 
Quantum technology is nowadays turning from the stage of experimental
setup design to that of quantum system engineering. The aim is to
produce tools for communication, information processing, and storage,
which rely on the principles of quantum mechanics, with the chance of
achieving much higher speeds and capacities than those of classical
devices. In this scenario, a new kind of quantum lab can be devised,
in which universality and programmability are crucial features, with
different tasks achieved by a basic set of devices.  \par A
``universal detector'' would allow the estimation of ensemble averages
of arbitrary operators using a single measuring apparatus, and by
changing only the data processing of the outcomes, according to which
ensemble average is estimated. Such a device would be very useful for
all kinds of quantum information processing tasks, such as in quantum
computation \cite{pop,nie}, teleportation \cite{ben,bra}, entanglement
detection \cite{wit}, and entanglement distillation protocols
\cite{ben2}. In some way the universal detector is similar to a
quantum tomographic apparatus \cite{tomo}: however, the latter would
typically require a {\em quorum} of observables---corresponding to a
set of devices or a single tunable device---whereas the universal
detector would just measure a fixed single observable on an extended
Hilbert space including an ancilla.  \par In this letter, we introduce
the general concept of universal detector, and characterize universal
detectors via a necessary and sufficient condition written in terms of
spanning sets of operators. We then show how such universal detectors
can be achieved via Bell measurements---i. e. measurement that are
described by projectors on maximally entangled states. The usefulness
of Bell measurements is not surprising. In fact, quantum
teleportation, dense coding, entanglement swapping \cite{ben,bra},
high-sensitivity measurements \cite{high}, tomography of quantum
operations \cite{tomochannel}, some types of quantum cryptography
\cite{cry}, and many other applications \cite{fabre,spettr,qlit},
require preparation of entangled states and/or Bell measurements.
However, entanglement is not an essential ingredient to build a
universal detector, since, as we will show in the following, the
universal observables of Ref. \cite{mauro} also enter the present
framework as a type of universal detector described by a POVM that is
based on local measurements and classical communication between system
and ancilla.  \par Let us start by defining the concept of universal
detector, or, more abstractly, of universal POVM. We are considering a
quantum system in a Hilbert space $\cal H$, coupled to an ancilla with
Hilbert space $\cal K$. A POVM $\{\Pi_i\}$, $\Pi_i\ge 0$ and
$\sum_i\Pi_i=I_{{\cal H}}\otimes I_{{\cal K}}$ on the Hilbert space
${\cal H}\otimes{\cal K}$ is {\em universal} for the system iff there
exists a state of the ancilla $\nu$ such that for any operator $O$ one
has
\begin{equation}
{\rm Tr} [\rho O]=
\sum_i f_i(\nu,O){\rm Tr}[(\rho\otimes\nu)\Pi_i]\,,
\label{def}
\end{equation}
where $f_i(\nu,O)$ is a suitable function of the outcome $i$ and the
operator $O$, which we will refer to as the {\em data processing}. The
detector will be named {\em universal} when it is described by a universal POVM. In order to give
a necessary and sufficient condition for universality, we need to
introduce some notation, and the concept of spanning set of
operators. We will use the following symbols for bipartite pure states
in ${\cal H}\otimes{\cal K}$ 
\begin{equation}
|A{\rangle\!\rangle}=\sum_{n=1}^{\hbox{\scriptsize dim}{\cal H}}
\sum_{m=1}^{\hbox{\scriptsize dim}{\cal K}} 
A_{nm}|n\rangle \otimes|m\rangle \;,\label{iso} 
\end{equation}
where $|n\rangle$ and $|m\rangle$ are fixed orthonormal bases for
$\cal H$ and $\cal K$, respectively. Equation \eqref{iso} exploits the
isomorphism \cite{bellobs} between the Hilbert space of the
Hilbert-Schmidt operators $A,B$ from ${\cal K}$ to ${\cal H}$, with
scalar product $\langle A,B\rangle =\hbox{Tr}[A^\dag B]$, and the
Hilbert space of bipartite vectors
$|A{\rangle\!\rangle},|B{\rangle\!\rangle}\in {\cal H}\otimes{\cal
K}$, with ${\langle\!\langle} A|B{\rangle\!\rangle}\equiv\langle A,B\rangle $. 
It is easy to show
the following identities
\begin{equation}
\begin{split}
&A\otimes B\dket{C}=\dket{ACB^T}\,,\\
&{\rm Tr}_{\cal K}[\dket{A}\dbra{B}]=AB^\dag\,,\\
&{\rm Tr}_{\cal H}[\dket{A}\dbra{B}]=A^TB^*\,,
\end{split}
\label{ids}
\end{equation}
where $T$ and $*$ denote transposition and complex conjugation 
with respect to the given bases, respectively. 
\par A spanning set for operators $A$ from ${\cal K}$ to ${\cal H}$
\cite{macca} is a set $\{\Xi_i \}$ that, along with its dual
$\{\Theta _i \}$, provides expansions for $A$ in the form
\begin{equation}
A=\sum _i {\rm Tr}[\Theta^\dag  _i A]\Xi _i\;.\label{sset}
\end{equation}
The completeness relation of the spanning set reads
\begin{equation}
\sum _i \langle \psi |\Xi _i |\phi \rangle \langle \varphi | \Theta
^\dag _i |\eta \rangle =\langle \psi |\eta \rangle \langle
\varphi|\phi \rangle \;, \label{ort}
\end{equation}
for any $\phi, \varphi \in {\cal H}$ and $\psi ,\eta \in {\cal
K}$. For continuous sets, the sums in Eqs. (\ref{sset}) and
(\ref{ort}) are replaced by integrals.   
\par Let us now consider a universal POVM on ${\cal H}\otimes{\cal
  K}$.  The elements $\{\Pi_i\}$ can be diagonalized as follows
\begin{equation}
\Pi_i=\sum_{j=1}^{r_i}\dket{\Psi^{(i)}_j}\dbra{\Psi^{(i)}_j}\,,
\label{diagp}
\end{equation}
where the vectors $\dket{\Psi^{(i)}_j}$ have norm equal to the $j$-th
eigenvalue of $\Pi_i$, and $r_i$ is the rank of $\Pi_i$. From 
the normalization condition $\sum _i \Pi _i= I_{\cal H}\otimes I_{\cal
  K}$, it follows that the set of operators $\{\Psi^{(i)}_j\}$ from
${\cal K}$ to ${\cal H}$ must be a spanning set itself.
\par The characterization of universal POVM's is then given by the
condition that there exists a state $\nu \in {\cal K}$ such that the
following operators
\begin{equation}
\Xi_i[\nu]\equiv\sum_{j=1}^{r_i}\Psi^{(i)}_j\nu^T\Psi^{(i)\dag}_j
\end{equation}
are a spanning set for operators on ${\cal H}$. In fact, using Eq. 
\eqref{diagp},  Eq. \eqref{def}  rewrites
\begin{equation}
{\rm Tr} [\rho O]=\sum_i f_i(\nu,O){\rm Tr}
\left[\rho\sum_{j=1}^{r_i}\Psi^{(i)}_j\nu^T\Psi^{(i)\dag}_j\right]\,,
\label{CNES}
\end{equation}
and this is true independently of $\rho$ iff
\begin{equation}
O=\sum_if_i(\nu,O)\Xi_i[\nu]\,.
\end{equation}
>From linearity one has
\begin{equation}
f_i(\nu,O)={\rm
  Tr}[\Theta^\dag_i[\nu]O] \;,\label{lin}
\end{equation}
where $\Theta_i[\nu]$ is the dual set of $\Xi_i[\nu]$. Hence, after
  finding a dual set for $\Xi_i[\nu]$, the data processing function is
  easily evaluated via Eq. (\ref{lin}).  \par We will now focus
  attention on Bell POVM's on ${\cal H}\otimes{\cal H}$.  In the
  notation of Eq. \eqref{iso}, maximally entangled vectors correspond
  to unitary operators \cite{bellobs}, and thus a Bell POVM has
  elements of the form
\begin{equation}
\Pi_i=\frac{\alpha_i}d\dket{U_i}\dbra{U_i}\,,
\end{equation}
where $d$ is the dimension of $\cal H$, $\alpha_i$ are suitable
positive constants and $U_i$ are unitaries.  When the POVM is
orthogonal, one has $\alpha _i =1$ and $\hbox{Tr}\left[U^\dag _i U _j
\right]=d\, \delta _{ij }$. Particular cases of Bell POVM's are those
in which $U_i$ are a unitary irreducible representation (UIR) of some
group ${\mathbf G}$. As an example, consider a projective UIR of an
abelian group, which therefore satisfies the relation
\begin{equation}
U _\alpha U_\beta U _\alpha ^\dag = e^{i c(\alpha ,\beta )} U_\beta \;.
\label{cab}
\end{equation}
In this case the Bell POVM is orthogonal, with number of elements
equal to the cardinality of the group $d^2$. One can show that a
suitable $\nu$ always exists such that the set of $\Xi_\alpha
[\nu]=\frac1d U_\alpha \nu^T U^\dag _\alpha $ is a spanning set. In
fact, for any $\nu $ such that $\mathrm{Tr}[U^\dag _\alpha \nu^T]\neq
0$ for all $\alpha $, using the identity
\begin{equation}
\sum _{\alpha  =1}^{d^2}e^{ic(\alpha ,\gamma )}e^{ic(\beta ,\alpha  )}=
d^2 \delta_{\gamma \beta }\;,
\end{equation}
one has 
\begin{equation}
\begin{split}
O&=\frac 1d \sum _{\beta =1}^{d^2} \mathrm {Tr}[U_\beta ^\dag O]U_\beta \\&= 
\frac {1}{d^3} 
\sum _{\beta ,\alpha ,\gamma =1}^{d^2} \frac 
{\mathrm {Tr}[U_\beta ^\dag O]}{\mathrm {Tr}[U_\beta ^\dag \nu ^T]}
{\mathrm {Tr}[U_\gamma  ^\dag \nu ^T]} \,U_\gamma \,e^{ic(\alpha ,\gamma )}
\,e^{ic(\beta ,\alpha )} \\& = 
\frac {1}{d^2} \sum _{\alpha  ,\beta =1}^{d^2} \frac 
{\mathrm {Tr}[U_\beta ^\dag O]}{\mathrm {Tr}[U_\beta ^\dag \nu ^T]}
\,e^{ic(\beta ,\alpha )} \,U _\alpha  \nu ^T U^\dag _\alpha \;.
\end{split}
\end{equation}
The dual set is then given by 
\begin{equation}
\Theta _\alpha  [\nu]=\frac {1}{d} \sum _{\beta  =1}^{d^2}\frac{U_\beta }
{\hbox{Tr}\left[U_\beta   \nu ^* \right]}\, e^{-ic(\beta  ,\alpha )} \;.\label{dua}
\end{equation} 
By identifying $U_1 \equiv I$, a possible choice of the ancilla
state is
\begin{equation}
\nu = \frac 1d I + \frac {1}{d (d^2-1)} \sum_{\alpha  > 1} U_\alpha \;.
\end{equation} 
\par For an explicit example, consider the group ${\mathbb
Z}_d\times{\mathbb Z}_d$ and its $d$-dimensional projective UIR
\begin{equation}
U_{m,n}=\sum _{k=0}^{d-1} e^{\frac {2\pi i}{d}km}|k \rangle \langle k
\oplus n|\;,\qquad
m,n\in [0,d-1]\;, 
\label{ZZ}
\end{equation}
which gives the Bell measurement used in the teleportation schemes of
Ref. \cite{ben}. The composition and orthogonality relations of the
set are given by 
\begin{eqnarray}
&&U_{m,n}\,U_{m',n'}\,U^\dag
  _{m,n}=e^{\frac{2\pi i}{d}(nm'-mn')}\,
  U_{m',n'}\;,  \label{propUs1} \\& & 
\hbox{Tr}\left[U^\dag_{p,q}\; U_{m,n}\right] = d\,\delta_{mp}\: \delta_{nq} 
\label{propUs2}\;.
\end{eqnarray}
Using Eqs. (\ref{lin}), (\ref{dua}), (\ref{propUs1}) and
(\ref{propUs2}), one easily evaluates the data processing function for
the operators $U_{p,q}$ in the form 
\begin{equation}
f_{m,n}(\nu,U_{p,q}) = 
\frac{\exp\left[
\frac{2\pi i}{d}(mq-np)\right]}{\hbox{Tr}[U^\dag _{p,q}\,\nu ^T]}\label{kunm}\;.
\end{equation}
\par As an example for the infinite dimensional case, consider the
displacement operators for a bosonic mode $a$, namely $D(\alpha
)=\exp(\alpha a^\dagger- \alpha ^* a)$, with $\alpha \in {\mathbb
C}$. Such operators are the elements of a projective UIR of the
Weyl-Heisenberg group, and generate the Bell measurement corresponding
to the continuous variables teleportation schemes of
Refs. \cite{bra,telep}. The composition and orthogonality relations
write 
\begin{eqnarray}
&&D(\alpha )D(\beta )D^\dag (\alpha )=e^{2i\hbox{\scriptsize Im}
(\alpha \beta ^*) }\,D(\beta )\;,\\ 
&&\hbox{Tr}[D^\dag (\alpha )D
(\beta )]= \pi\,\delta ^{(2)}(\alpha -\beta )\;,
\end{eqnarray}
where $\delta ^{(2)}(\alpha )\equiv (1/ \pi ^2)\int_{\mathbb C}d^2 \gamma \,
e^{\alpha \gamma ^*-\alpha ^* \gamma }$ denotes the Dirac-delta
on the complex plane.  The processing function is given by 
\begin{equation}
f_\alpha (\nu, O)=
\int_{\mathbb C}\frac{d^2 \beta }{\pi }\,e^{ \alpha ^*\beta  -\alpha 
  \beta ^* }\frac{\hbox{Tr}[D^\dag (\beta ) O]}{\hbox{Tr}
[D^\dag (\beta ) \nu ^\tau]}
\;.\label{int}
\end{equation}
The Dirac vectors $|D(z) \rangle \!\rangle$ are the eigenvectors
with eigenvalue $z$ of the ``current'' $Z=a-b^\dag $, which in the
case of e. m. radiation is the Bell observable of heterodyne \cite{unc},
eight-port homodyne \cite{ott,leo} or six-port homodyne detectors
\cite{tri}, whereas for atoms coupled with two light fields the observable
is achieved by measuring the corresponding phase-shifts \cite{wil}.  
The present infinite dimensional case, however, needs care in checking
convergence of the integral in  Eq. (\ref{int}).  For example, if we
take the vacuum state $\nu=|0\rangle\langle 0|$, the universal
measurement will be the phase-space averaging with the so-called Q-function
$Q(z)=\frac{1}{\pi}\langle z|\rho|z\rangle$ ($|z\rangle$ coherent
state), and we know that this gives expectations only for
operators admitting anti-normal ordered field expansion \cite{bal}.
In particular, the matrix elements of the density 
operator cannot be recovered in this way \cite{tomo}. Therefore, in
infinite dimensions the universality can be limited by convergence.
\par There are universal Bell POVM's also from non-abelian groups. For
example, the $SU(2)$ group corresponds to a non orthogonal POVM
\cite{telep}. Consider the $j$-dimensional UIR of 
the $SU(2)$ group \cite{perelom}, parameterized as $U(\psi , \vec n)=
\exp(i\psi \vec J\cdot\vec n)$, where $\psi\in[0,2\pi)$, $\vec n =(\sin\theta
\cos\varphi,\sin\theta \sin\varphi, \cos\theta)$ is a unit vector 
on a sphere $S^2$, and $J_{\alpha}$ are customary
angular momentum operators.
The projectors on maximally entangled states
\begin{equation}
\Pi (\psi, \vec n)= |U(\psi, \vec n)
\rangle \!\rangle
\langle \!\langle U(\psi, \vec n) | \;\label{psu}
\end{equation}
provide the resolution of the identity\cite{telep}
\begin{equation}
\frac{2j+1}{4\pi ^2}
\int _0 ^{2\pi }d\psi \,\sin^2 \frac \psi 2 
\int _{S^2}d\vec n \,
\Pi(\psi, \vec n)=I _{\cal
  H}\otimes I_{\cal H}\;.
\end{equation}
Notice, however, that the states $|U(\psi, \vec n) \rangle \!\rangle$
are not orthogonal, namely the set is overcomplete for ${\cal
H}\otimes{\cal H}$.  The universality of this POVM can be proved by
invoking the fact that the projectors on spin coherent states $|\psi,
\varphi;m\rangle$ are also an operator spanning set.  In fact, spin
coherent states are obtained by applying the unitary operators
\begin{equation}
D(\psi, \varphi)=e^{i\frac \psi 2 (J_+e^{-i\varphi} +J_-e^{i\varphi
})}\;\label{dpsi}
\end{equation}
on a fixed eigenstate $|m \rangle $ of $J_z$, namely $|\psi,
\varphi;m\rangle\doteq D(\psi, \varphi)|m \rangle$. Notice that the
operators in Eq. (\ref{dpsi}) {\em do not} form a group, nevertheless
one has the completeness relation
\begin{equation}
\frac{2j+1}{4\pi }
\int _0 ^{2\pi }d\psi \int _0 ^{\pi }d\varphi\,\sin(\varphi )
|\psi, \varphi;m\rangle\langle\psi, \varphi;m|= I _{\cal H}\;,
\end{equation}
and the $P$-representation for any operator $O$
\begin{equation}
\begin{split}
O=\frac{2j+1}{4\pi }\int _0 ^{2\pi }d\psi \int _0 ^{\pi
}d\varphi\,\sin(\varphi )\times\\ P_O(\psi, \varphi)\,|\psi,
\varphi;m\rangle \langle\psi, \varphi;m|\;.\label{pfun}
\end{split}
\end{equation}
Explicit constructions for the function $P_O(\psi, \varphi)$ can be
found in Ref. \cite{perelom}.
The above expansion is similar to the P-function representation of
quantum optics, however, here the function $P_O(\psi, \varphi)$ is
always well defined, due to the finite dimensionality of the Hilbert
space.  The universality of the Bell POVM in Eq. (\ref{psu}) is now
proved by showing that there exists a state $\nu $ such that $U(\psi, \vec n)
\nu ^T U^\dag (\psi, \vec n)$ is a spanning set. In fact, using the following
$SU(2)$ change of parameterization 
\begin{equation}
U(\psi, \vec n)= D(\psi ', \varphi ')\,e^{2i \theta ' J_z}\;,\label{fact}
\end{equation}
according to Eq.  (\ref{pfun}), for any $\nu ^T=\sum
_{m=-j}^j p_m \,|m \rangle \langle  m|$,  the following operators make
a spanning set \cite{unpub}
\begin{equation}
U(\psi, \vec n) \nu^T U^\dag (\psi, \vec n) = D(\psi ', \varphi ') \,\nu ^T\,
D^\dag (\psi ', \varphi ')\;.
\end{equation}
\par Another interesting example is represented by the group $SU(d)$. In
this case the universality of the corresponding Bell POVM is proved by
constructing the dual set of $\Xi_\alpha[\nu]=
U_\alpha\,\nu ^T\,U^\dag_\alpha$. A dual set for any state 
$\nu$ on $\cal H$ is given by
\begin{equation}
\Theta_\alpha[\nu ]=U_\alpha \xi U^\dag_\alpha\;, 
\end{equation}
for arbitrary $\xi $ such that $\hbox{Tr}[\xi ]=d$ and $\hbox{Tr}[\nu
^T \xi^\dag ]=d^2$. For example, for pure $\nu ^T =|\phi \rangle \langle
\phi |$, one can take 
\begin{equation}
\xi =\frac {d}{1-F}\left[(d-F)|\phi\rangle\langle\phi| - 
(d-1)|\psi\rangle\langle\psi|\right ]\;,
\end{equation}
with any state $|\psi \rangle $ with $F\equiv
|\langle\psi|\phi\rangle |^2< 1$.  The corresponding data processing function is
\begin{eqnarray}
&&
f_\alpha(|\phi\rangle\langle\phi|,O)= 
\frac {d}{1-F} [(d-F)\langle\phi|U^\dag_\alpha
OU_\alpha|\phi\rangle \nonumber \\& & - (d-1)\,\langle\psi|U^\dag_\alpha
OU_\alpha|\psi\rangle ]\,.
\end{eqnarray}
\par All previous examples presented universal POVM's which are Bell
measurements.  However, by enlarging the ancillary Hilbert space, one
can obtain separable POVM's that are universal. The following example
was first introduced in Ref. \cite{mauro}.  Let us consider a spanning
set $\{C(l)\ \;,l=1,2,...,L\}$, for operators on $\cal H$ such that
all $C(l)$ are normal, namely they have orthogonal eigenvectors
$|c_k(l)\rangle $, and
\begin{equation}
C(l)=\sum_k c_k(l) |c_k(l)\rangle\langle c_k(l)|\;.
\end{equation}
 Notice that necessarily one has $L \geq (\hbox{dim}({\cal H}))^2$. By
taking an ancillary Hilbert space $\cal K$, with $\hbox{dim}({\cal
K})=L$, and an orthonormal basis $\{|l \rangle \}$ one can write the
following orthogonal POVM for ${\cal H}\otimes {\cal K}$
\begin{equation}
\Pi _{k,l}= |c_k(l)\rangle \langle c_k(l) |\otimes |l\rangle \langle l
| \;,\label{loc}
\end{equation}
which can be achieved by local measurement and classical
communication between system and ancilla. The universality of
$\Pi_{k,l}$ easily follows by using Eq. (\ref{def}) with the
processing function 
\begin{equation}
f_{k,l}(\nu ,O)=\frac{\hbox{Tr}[C^\dag
      (l)0]}{\langle l|\nu |l \rangle }c_k(l)\;,
\end{equation}
with the condition $\langle l|\nu |l \rangle \neq 0$ for all $l$.  
\par The form of the above separable universal POVM's opens some
questions on the general structure of universal POVM's. For example,
it is possible that also universal Bell POVM's could be constructed
using general unitary spanning sets that are not a group
representation, and the role of such symmetry is probably
not essential. Also, it is likely that when the ancilla space has the  
same dimension of the system, then the universal POVM must be
Bell. Moreover, a general classification of universal POVM's along
with the pertaining ancilla states and data processing functions is
needed for performance optimization.  
\par This work has been sponsored by INFM through the project
PRA-2002-CLON, and by EEC through the ATESIT project IST-2000-29681.
G. M. D. also acknowledges partial support from Department 
of Defense Multidisciplinary University Research Initiative (MURI)
program administered by the Army Research Office under Grant 
No. DAAD19-00-1-0177.

\end{document}